\documentclass[conference,compsoc]{IEEEtran}
\IEEEoverridecommandlockouts

\def\BibTeX{{\rm B\kern-.05em{\sc i\kern-.025em b}\kern-.08emT\kern-.1667em\lower.7ex\hbox{E}\kern-.125emX}}

\usepackage[hidelinks]{hyperref}
\usepackage{tikz}
\usepackage{soul}
\usepackage{xspace}
\usepackage{amsmath}
\usepackage{cleveref}
\usepackage{etoolbox}
\usepackage{pifont}
\usepackage{booktabs}
\usepackage{xcolor} 
\usepackage{enumitem}
\usepackage{calc}

\newtoggle{draftmode}
\settoggle{draftmode}{true}

\newcommand\graphsize{0.8}
\newcommand\tblvsp{0pt}
\def\etal{\emph{et al}.\xspace}

\definecolor{Newstuff}{RGB}{70, 104, 163}
\definecolor{Comment}{RGB}{204,102,0}
\definecolor{Delete}{RGB}{171, 56, 56}

\iftoggle{draftmode} {
	\newcommand{\del}[1]{{\color{Delete}#1}}
    \newcommand{\todo}[1]{\marginpar{\textcolor{Comment}{TODO\\\footnotesize #1}}}

} {
	\newcommand{\del}[1]{}
	\newcommand{\todo}[1]{}
  
}

\newcommand{\sysname}{AnonSat\xspace}

\newcommand{\adv}{$\mathcal{A}$\xspace}
\newcommand{\maxhops}{$max\_hops$\xspace}
\newcommand{\timeout}{$gateway\_timeout$\xspace}

\begin{document}
\date{}

\title{Don't Shoot the Messenger:\\Localization Prevention of Satellite Internet Users}

\author{
\IEEEauthorblockN{David Koisser}
\IEEEauthorblockA{Technical University of Darmstadt\\
david.koisser@trust.tu-darmstadt.de\\~\\}

\IEEEauthorblockN{Marco Chilese}
\IEEEauthorblockA{Technical University of Darmstadt\\
marco.chilese@trust.tu-darmstadt.de}
\and
\IEEEauthorblockN{Richard Mitev}
\IEEEauthorblockA{Technical University of Darmstadt\\
richard.mitev@trust.tu-darmstadt.de\\~\\}

\IEEEauthorblockN{Ahmad-Reza Sadeghi}
\IEEEauthorblockA{Technical University of Darmstadt\\
ahmad.sadeghi@trust.tu-darmstadt.de}
}

\maketitle
\pagestyle{plain}


\begin{abstract}
Satellite Internet plays an increasingly important role in geopolitical conflicts.
This notion was affirmed in the Ukrainian conflict escalating at the beginning of 2022, with the large-scale deployment of the Starlink satellite Internet service which consequently demonstrated the strategic importance of a free flow of information.
Aside from military use, many citizens publish sensitive information on social media platforms to influence the public narrative.
However, the use of satellite communication has proven to be dangerous, as the signals can be monitored by other satellites and used to triangulate the source on the ground.
Unfortunately, the targeted killings of journalists have shown this threat to be effective.
While the increasing deployment of satellite Internet systems gives citizens an unprecedented mouthpiece in conflicts, protecting them against localization is an unaddressed problem.

To address this threat, we present \sysname, a novel scheme to protect satellite Internet users from triangulation.
\sysname works with cheap off-the-shelf devices, leveraging long-range wireless communication to span a local network among satellite base stations.
This allows rerouting users' communication to other satellite base stations, some distance away from each user, thus, preventing their localization.
\sysname is designed for easy deployment and usability, which we demonstrate with a prototype implementation.
Our large-scale network simulations using real-world data sets show the effectiveness of \sysname in various practical settings.

\end{abstract}


\section{Introduction}
\label{sec:intro}
The Internet has fundamentally changed the way conflicts unfold and are perceived on the global stage.
A crucial aspect is the growing role of social media with civilians sharing, publishing, and forwarding information, including sensitive strategic data.
The term \emph{hybrid warfare}~\cite{Ihybridwar} alludes to the increasing focus on information warfare, such as disinformation campaigns.
For example, allegedly, journalists were specifically and lethally targeted in both the second Chechen war and the Syrian civil war, to control the public narrative~\cite{Ikillmessenger}.
Today, individual citizens can leverage social media to reach a global audience, further escalating the dynamics of (dis-)information.

This new paradigm is particularly visible in the Ukrainian conflict that escalated in February 2022.
United Nations appointed independent rights experts warned that journalists in Ukraine are targeted and in danger~\cite{Iunjournalists}, with 15 journalists confirmed to have been killed in Ukraine in 2022~\cite{Ijournalistskilled}.
Another perspective on the conflict claims Ukraine and its citizens got the upper hand in the so-called \emph{social media war}~\cite{Isocialmediawar}.
Indeed, the U.S. government recognized the importance of a free flow of information among Ukrainians.
After spending millions of dollars to fund the widespread deployment of Starlink satellite Internet terminals and service in Ukraine~\cite{Iusgovpaysstarlink}, the service quickly reached over 150\,000 users shortly after deployment~\cite{I150kusers}.
While a large share of Starlink's usage in Ukraine seems to be military, the Starlink smartphone app was downloaded over 806\,000 from Ukraine, making it the most downloaded app at the time~\cite{Istarlinkapptop}.
As a response, Russia has started a barrage of cyber and jamming attacks on Starlink since~\cite{Istarlinkattacks}, yet they were repelled~\cite{Istarlinkupdate}.
While the precedent of Starlink's satellite Internet in a conflict is still ongoing at the time of writing, the E.U. already announced a deal to deploy its own satellite Internet system~\cite{Ieustarlink}.

Clearly, satellite-based Internet has a significant impact, as citizens and journalists gain the ability to freely share, e.g., sensitive information, evidence of war crimes, or timely warnings.
However, openly sharing information also carries great risks, and thus, many social media companies have added additional security measures to protect Ukrainian users, along with some guidelines to minimize risks~\cite{Istaysafe}.
Moreover, there is a specific danger when using satellite-based communication, especially when used to publish sensitive information.
As satellite signals can be monitored by virtually anyone in the sky and space above, satellite uplink communications can also be used to geolocate their users on the ground by triangulating their signals.
One example is the killings of two American journalists~\cite{Ijournaliststriked} and another a missile strike on the leader of the Chechen republic~\cite{Ichechenstrike}.
In both cases it is assumed that the attacks were possible by tracing satellite phones.
While official confirmation of such state-backed attacks is rare, it was shown that techniques to geolocate transmitters by satellites (target tracking) are practical~\cite{elgamoudi2021survey}.
After a security researcher gained traction with a tweet warning Ukrainian Starlink users potentially being geolocated and becoming targets~\cite{Itweet}, the CEO of Starlink's company issued a public warning to Ukrainian Starlink users~\cite{Imuskwarning}.

Considering the scale at which a satellite-based Internet can be monitored, and worse, individual users triangulated to get their physical position, it is an important aspect to protect citizens against this threat while preserving this novel and free flow of information during conflicts.
However, there are no works to properly address this issue in a practical manner.
On the one hand, using typical Internet encryption (i.e., TLS) on the communication channel is insufficient.
Eavesdropping on satellite communication targets the actual physical medium, whereas TLS is a high-level protocol not designed to provide anonymity.
The Tor network aims to fix this problem for the traditional Internet.
However, our case has a crucial difference, as satellite communications can be monitored by any satellite and, worse, triangulated to geolocate the user.
For example, numerous attacks on Tor assume an adversary can do \emph{entry point} monitoring~\cite{murdoch2007sampled,yang2017active}.
While typically an ambitious position for the adversary, with a satellite-based Internet, it becomes quite straightforward, as the connection is first sent to the satellite before reaching the Tor network.
Other attacks on Tor assume the adversary can monitor both the \emph{entry} and \emph{exit point}~\cite{bauer2009predicting,le2011one,palmieri2019distributed}.
However, if the adversary aims to prevent a user from publicly sharing information, it may anticipate and monitor popular social media sites, such as Twitter, for the exit point.

Prior works on \emph{location privacy} in mesh and wireless sensor networks have different shortcomings~\cite{kamat2005enhancing, xi2006preserving, shaikh2010achieving, li2009preserving, hong2005effective, kamat2009temporal, rios2011exploiting, el2010hyberloc, fan2009efficient, fan2010preventing, yang2013towards, mehta2007location, ouyang2006entrapping, kazatzopoulos2006ihide, wang2009privacy, shao2009cross, zhang2006arsa, sun2010sat, misra2006efficient, luo2010location}.
For example, some works have impractical assumptions for our purposes and others induce significant overheads.
There are also works that aim to establish a reliable network in case the existing infrastructure fails, called \emph{emergency networks}~\cite{portmann2008wireless,zhao2019uav,panda2019design,deruyck2018designing,pan2021uav,lin2021adaptive}.
These approaches are typically based on specialized hardware, such as vehicles equipped with bulky communication equipment and even flying vehicles, and custom network protocols to facilitate basic communication, making them quite impractical for our purposes.
There are also emergency networks working with satellites~\cite{zhou2021integrated,iapichino2008advanced,patricelli2009disaster}; yet, they do not consider protection against triangulation.
We will discuss the related work more thoroughly in \Cref{sec:rw}.
\\~\\
\indent
In summary, the remote monitoring and the possibility to triangulate satellite Internet users is a global threat to the new-found free flow of information by citizens.
To the best of our knowledge, there is no existing system that prevents geolocating satellite Internet users.
In this paper, we present \emph{\sysname} to close this gap.
\\~\\
\noindent\textbf{Goals \& Contributions:}
Our primary goal is to hide the geographic position of satellite Internet users in case their connection is being triangulated.
\sysname works by leveraging long-range wireless communication to span a simple network among satellite base stations.
Our system is agnostic with respect to the used wireless communication technology.
Leveraging this local network, a client's WAN connection is routed to another randomly selected satellite base station, which acts as a delegate to do the actual connection uplink to the satellite.
Further, the targeted satellite base station is regularly changed to avoid tracing back a long-lasting connection.

Our secondary goal is to focus on the accessibility of our system.
Therefore, \sysname is designed to work with cheap and simple devices, and thus, it can be deployed with widely available hardware and does not require impractical extensions on the user's device, such as a specific app or radio device.
Further, we aim at the usage of popular Internet services, like Twitter or WhatsApp, refraining from custom network protocols.
\sysname effectively protects users from being geolocated and becoming targeted.

\vspace{1em}
\noindent Our main contributions include:

\begin{itemize}
	\item \sysname is the first scheme to address the triangulation of satellite Internet users by rerouting connections to more distant satellite base stations.
	We introduce two security parameters that adjust the selection of routes to avoid geolocating a user over time.
	\item We derive requirements for wireless communication technologies and give an overview of possible candidates that can be used to enable \sysname.
	Similarly, due to our aim to design a practical system, we discuss numerous approaches, such that \sysname can access typical Internet services without needing to install custom software or hardware on the user's device.
	\item We implemented a proof of concept demonstrating the feasibility of \sysname, leveraging a cheap Raspberry Pi equipped with a LoRa shield for the local network.
	\item We further developed a large-scale simulation using real-world data sets to evaluate key aspects of \sysname in different environments, such as effective distances from the user or the use of more powerful wireless technologies.
\end{itemize}


\section{System Model}
\label{sec:system_model}

Our system model consists of the following entities:
	A \textbf{network} is a collection of \emph{gateways}, \emph{clients}, and satellites providing internet access.
    Each \textbf{gateway} is equipped with a base station, i.e., means of providing access to the Internet via satellite communication, a radio transmitter to connect to the \emph{local network}, and two WiFi access points to provide access to the \emph{WiFi Network}.    
    A \textbf{local network} is spanned among the \emph{gateways} via their radio transmitter capable of communicating with each other.
    A \textbf{WiFi network} is a direct connection between \emph{clients} and \emph{gateways} separated into two WiFi access points.
    One is a connection to the gateway's WAN, directly uplinking to the satellite internet.
    The other provides a secure WiFi connection via our \sysname system.    
    \textbf{Clients} are simple end-user devices, such as smartphones, which aim to establish a secured WAN connection to send sensitive data.
    For this, \emph{clients} simply connect to the secure \emph{WiFi Network} provided by a close-by \emph{gateway}.

Note, for simplicity, we assume all \emph{gateways} have a base station providing Internet access, even though in a real-world scenario simple relay nodes for the \emph{local network} may also be deployed.
We further assume that standard means of Internet access are impaired or even unavailable entirely, and thus clients need to rely on satellite Internet.

Furthermore, \emph{gateways} are equipped with certificates to identify each other and to establish secret key pairs to enable symmetric encryption between any two \emph{gateways}.
We assume the \emph{gateways} can trust each other's certificates.
For example, in the real world, this could be realized via exchanging certificates via direct contacts in combination with a Web-of-Trust approach.

\subsection{Adversary Model and Assumptions}
\label{sec:adv_model}
The adversary \adv has the goal of geolocating a specific client.
To do this, \adv has a range of satellites deployed, which can eavesdrop on the Internet communication between base stations and the receiving satellite.
We assume that \adv is able to correlate the communication data to identify a specific client.
Further, \adv is capable of triangulating the sending base station via the mentioned satellites, which was shown to be practical~\cite{elgamoudi2021survey}.
Thus, as the client has to use the closest base station via a close-range WiFi connection, \adv can infer the client's approximate geographic position.
However, as our system aims to reroute the client's connection to another gateway, simply triangulating the base station is not enough to geolocate the client.

We assume \adv is not capable of establishing a holistic view of the local network.
To achieve this, \adv would need to monitor a significant number of local network connections, which also requires prolonged physical proximity in a multitude of locations.
This contradicts the scenario we are targeting, i.e., an active conflict zone, as widespread deployment of eavesdropping devices is infeasible.
However, \adv is able to intercept individual messages sent between gateways.
This assumption is comparable to the Tor network, which can also be broken by an adversary with a global view of the network; yet, in practice, this is hard to achieve.
We further argue due to the nature of the limited range of each node (discussed in \Cref{sec:tech:lpwan}), the local network cannot be surveilled by satellites to attain a global view.

We will thoroughly discuss several possible local attacks in \Cref{sec:security}, including jamming attacks that deal with similar assumptions.
Nevertheless, our system focuses on preventing remote and globally applicable triangulation via satellites.

Further, we assume the WiFi connection between the client and gateway cannot be intercepted by \adv due to its close-range nature.
We also assume \adv aims to minimize any collateral damage, as this might have grave political repercussions~\cite{Isanctions,Iiccwarrant}.
Finally, we further assume \adv cannot forge digital signatures or break symmetric encryption.

\subsection{Requirements}
\label{sec:requirements}
To formalize the setting outlined in the Introduction, we aim to design a secure satellite Internet scheme with the following requirements:
\begin{enumerate}
	\item[R.1] \label{req1:baseloc}\emph{Prevent geolocating base station:} 
	The scheme shall prevent \adv from geolocating the client.
	More specifically, the gateway uplinking traffic to the WAN shall not indicate the geographic position of the client.
	For example, as a client has to use the closest gateway, if this gateway uplinks the client's traffic to the WAN, \adv may assume the client is very close.
	\item[R.2] \label{req2:localloc}\emph{Prevent local geolocation leakage}:
	In addition to requirement~\hyperref[req1:baseloc]{R.1}, \adv may intercept individual messages in the local network traffic and trace back the actually used gateway by the client.
	Thus, the scheme shall further prevent geolocation leakage in terms of the local network.
	\item[R.3] \label{req3:compatible}\emph{Internet compatibility:}
	The use of, e.g., simplified network protocols may greatly increase the performance of the scheme.
	However, this implies that most common Internet services are not accessible, and thus, the scheme shall be compatible with most Internet services.
	\item[R.4] \label{req4:minreq}\emph{Out-of-the-box for clients:} 
	From the client's point of view, the scheme shall impose minimal requirements on the client.
	For example, a client may need to unexpectedly and urgently send some sensitive data using a smartphone.
	In such a case, requiring the client to install an additional app or even an additional hardware device is impractical.
\end{enumerate}


\section{\sysname Design}
\label{sec:overview}
\label{sec:theory}
In this section, we focus on the general design of \sysname.
However, as our focus is designing a practical system (cf. \Cref{sec:requirements}), we will discuss essential technical aspects in \Cref{sec:tech}.
\begin{figure}[t]
	\centering
	\includegraphics[width=\graphsize\columnwidth,trim={0 0 0 0},clip]{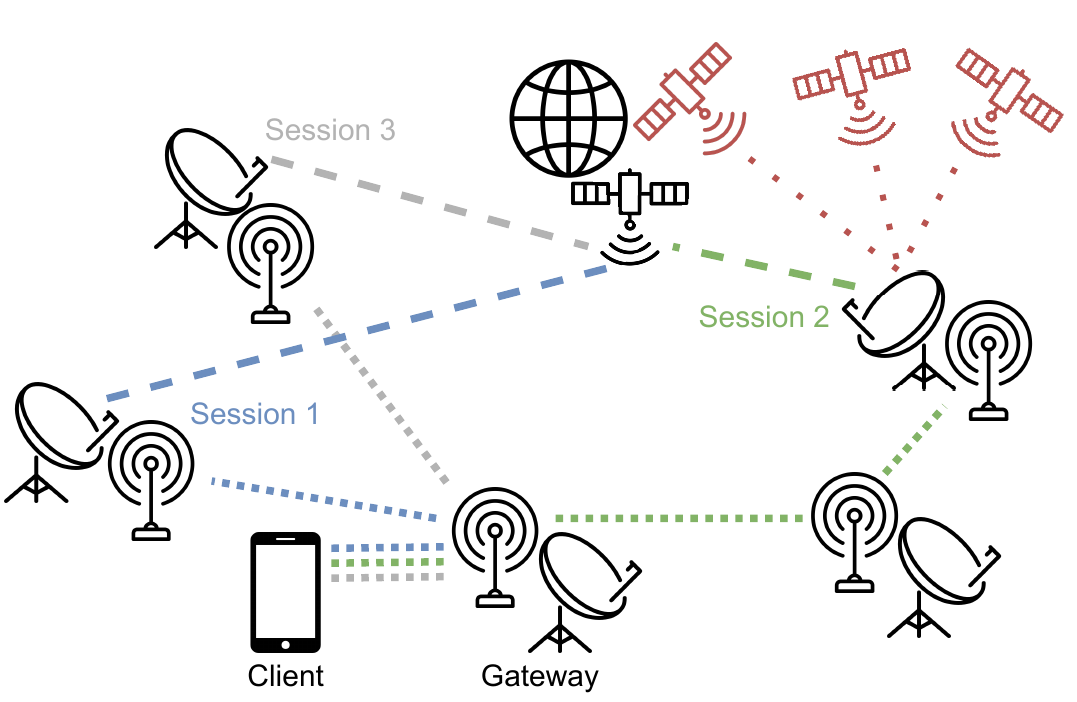}
	\caption{An example setup of \sysname with five gateways and one client.} 
	\label{fig:overview}
\end{figure}
\Cref{fig:overview} illustrates a simplified setup of \sysname with five gateways, each with a base station to connect to the Internet and a local radio transmitter to communicate with other gateways.
The client uses a smartphone to upload sensitive data to a gateway in close proximity, called the \emph{origin}.
For example, the client may try to publish an incriminating picture on Twitter.
Instead of directly forwarding the data over the satellite link, the gateway randomly selects an output gateway.
In \emph{Session 1} in the figure, the origin will then transmit the client's data over the local radio connection to the output gateway, which in turn will do the actual satellite transmission.
Thus, if the output gateway is identified and triangulated by \adv, the actual geographic position of the client stays hidden.
After a chosen amount of time, the origin gateway will select a new output gateway for this client's connection in \emph{Session 2}.
This new output gateway is only reachable via an intermediate gateway, and thus, establishes a 2-hop connection.
As seen in \emph{Session 3}, the origin will continue to change the client's output gateway regularly while the connection lasts.

\paragraph{\emph{Changing Gateways}}
While we assume \adv cannot eavesdrop on the entire local gateway network (cf. \Cref{sec:adv_model}), \adv may be able to intercept individual messages.
As a first step, all communication between the gateways is per-hop encrypted.
Thus, a forwarding gateway will receive an encrypted message, decrypt it, encrypt it with the key established with the next-hop gateway, and forward it.
However, if \adv intercepts enough messages in relation to a specific client, \adv may be able to trace the origin gateway, and thus, geolocate the client eventually.
Therefore, we introduce the security parameter \timeout, which defines a timeframe.
When a client starts a connection and the origin gateway selects a random output gateway, a timer is started.
After \timeout time, the origin will select a new output gateway.
Choosing a proper value for \timeout depends on the bandwidth and delay of the local gateway network.
Note that \timeout may also define a message count instead of a timeframe.
If \timeout is set too high, \adv has an increasing chance to trace back the origin.
Setting \timeout too low may lead to technical problems with the client's connection.
We will discuss this further in \Cref{sec:tech:routes}.
Note, \sysname's goal is not to make all traffic indistinguishable from unobjectionable traffic, as current approaches, such as Dummy Data Sources create non-negligible overhead (we discuss this in \Cref{sec:rw}).

\paragraph{\emph{Distance to Origin}}
The selection of the output gateways has some crucial implications as well.
If the origin only selects very close gateways, then the clients are not adequately protected.
If the output gateways are too far away, this would require many hops between the origin and the output gateways, negatively affecting the performance of \sysname.
Thus, we introduce the security parameter \maxhops, which defines how many hops away the output gateway shall be from the origin.
Increasing this parameter increases the average distance from the client (i.e., the origin gateway) to the output gateway, and thus, increasing the protection of the client.
However, increasing \maxhops will negatively affect the performance of the client's connection.

\paragraph{\emph{Output Selection Bias}}
While simply routing the client's traffic to output gateways \maxhops hops away is fine for individual short sessions, we need to consider a crucial aspect for longer sessions.
Suppose we have a uniformly spread network of gateways and a client that needs a long-lived connection to, e.g., upload many pictures.
In this case, many gateway changes happen over time and the selected gateways will eventually be selected in all directions from the origin.
Simply put, \adv can observe these changes and will be able to draw a circle containing all output gateways.
The gateway closest to the centroid of this circle is most likely the origin gateway; thus, endangering the client.
To counteract this effect, we additionally introduce a selection bias for the output gateway.
For each client, the origin gateway will generate a random \emph{direction} and \emph{weight} bias. 
When selecting a new output gateway, the origin will prefer random output gateways in the given direction and with the assigned weight.
In a practical context, the direction can simply be a bias for selecting the next neighbor gateway via an index without the need to consider the gateways' geolocations.
This way, the centroid of the mentioned circle shifts to a random direction, and thus, cannot be used to trace the origin.
The selected biases will be preserved for each client.

Additionally, this allows us to change the role of the \maxhops security parameter.
Instead of always selecting an output gateway exactly \maxhops hops away, we can select a range of hops between 0 and \maxhops.
This does not weaken the previously discussed selection bias.
However, the range improves the client's network performance on average, as some output gateways may be only one hop away and fewer hops mean better performance for the client.
Note that it is important for the origin to select itself as the output gateway.
Otherwise, \adv can simply identify the origin by checking which gateway never sends.


\section{Technical Considerations}
\label{sec:tech}
After we described the theoretical design in \Cref{sec:theory}, it is crucial to consider the technical challenges of \sysname to satisfy requirements \hyperref[req3:compatible]{R.3} and \hyperref[req4:minreq]{R.4}.
Therefore, this section will discuss key practical aspects to implement our theoretical design.
Note, we focus on available techniques or techniques currently in deployment, i.e., we will not address recent research approaches, as these might take many more years until ready for deployment.
Our focus is on techniques usable in a practical deployment now or in the near future.

\subsection{Wireless Communication Technology}
\label{sec:tech:lpwan}
As \sysname relies on a local network between gateways, this section discusses possible options for wireless communication technologies.
Fortunately, due to the prevalence of the Internet of Things (IoT), there have been many proposals and advancements in recent years to enable remote IoT devices to connect directly to the Internet or via a mesh network.
We will leverage these advancements for \sysname.

\setlength{\tabcolsep}{3.6pt}
\begin{table}[t]
    \centering
    \caption{Overview of long-range radio transmission technologies applicable to \sysname.}
    \label{tbl:lpwans}
    \begin{tabular}{llllc}
        \toprule
        Name & \parbox{\widthof{Maximum}}{Maximum\\Data\\Rate} & \parbox{\widthof{Range}}{Range\\(urban)} & \parbox{\widthof{Range}}{Range\\(rural)} & \parbox{\widthof{Unlicensed}}{Unlicensed\\Frequency\\Bands} \\ 
        \midrule

        \vspace{0.5em}
        LoRa Sub-GHz~\cite{foubert2020long} & \parbox{\widthof{50\,kbps (FSK)}}{27\,kbps\\50\,kbps (FSK)} &  5\,km & 15\,km & \ding{51} \\ \vspace{0.5em}
        LoRa 2.4\,GHz & \parbox{\widthof{1\,Mbps (FLRC)}}{250\,kbps\\1\,Mbps (FLRC)} & 1\,km & n/a & \ding{51} \\ \vspace{0.5em}
        LTE-M Cat-M1~\cite{ltem-aws} & \parbox{\widthof{4\,Mbps (Cat-M2)}}{1\,Mbps\\4\,Mbps (Cat-M2)} & 1\,km & 10\,km & \ding{55} \\ 
        NB-IoT Cat-NB2~\cite{mekki2019comparative} & 200\,kbps & 1\,km & 10\,km & \ding{55} \\
        DASH7~\cite{foubert2020long} & 166\,kbps & 5\,km & n/a & \ding{51} \\
        Weightless-W~\cite{foubert2020long} & 10\,Mbps & 5\,km & n/a & \ding{55} \\
        \bottomrule
    \end{tabular}
    \vspace{\tblvsp}
\end{table}
\setlength{\tabcolsep}{6pt}

\Cref{tbl:lpwans} shows an overview of the technologies we considered.
Note, this table is not comprehensive of all available technologies, as we carefully selected technologies we deem applicable to \sysname.
For example, while Sigfox is a mature wireless technology already deployed in many regions around the world, it only supports 100\,bps data rates~\cite{mekki2019comparative}, which is too slow to be meaningfully used to connect to traditional Internet services (violating requirement~\hyperref[req3:compatible]{R.3}).

One may also consider repurposing existing infrastructure, such as the cellular network for mobile Internet.
However, in a conflict, existing infrastructure (e.g., cell towers) is unreliable or may not be functional at all, making satellite Internet necessary in the first place.
In addition, repurposing this infrastructure for a disaster network is not feasible as cellular networks do not have mesh network capabilities and access to the hardware is limited.

For the \emph{Maximum Data Rate}, we also considered alternatives, like the Fast Long Range Communication (FLRC) for Lora 2.4\,GHz, which uses demodulation and error correction techniques for a significantly improved data rate~\cite{lora24chip}.
Further, the last column of \Cref{tbl:lpwans} shows if the respective technology uses unlicensed frequency bands.
For example, Lora supports a variety of different sub-GHz bands, which are publicly usable in the respective region, e.g., North America allocates different bands than Europe~\cite{mekki2019comparative}.
While the other technologies use licensed bands, we expect that legal restrictions play a lesser role in the settings applicable to \sysname or may even be officially lifted. 

\subsection{MTU Size Mismatch}
\label{sec:tech:mtu}
One technical hurdle with a significant practical impact is the Maximum Transmission Unit (MTU), which defines the maximum amount of bytes sent in a single network layer transaction.
Considering our scenario targeting the Internet, typically the MTU of Ethernet is used (1500\,bytes).
The hurdle arises when using a wireless communication technology, which only supports a smaller MTU.
For example, LoRa restricts the MTU to 255\,bytes, which leads to problems with Internet servers expecting a large MTU, such as extra negotiation steps for smaller messages and spurious retransmissions.
For example, in our preliminary tests using a small MTU, we saw significant delays with TLS handshakes, even to a point in which some servers simply declined the session entirely.
Another downside of a small MTU is the overhead of the headers, such as the IP and TCP protocols.
With a small MTU, the headers consume a significant share of the bandwidth.

Therefore, a mechanism is required to split Ethernet MTUs received by the client to fit them into, e.g., multiple LoRa frames for forwarding them over the local network.
These kinds of operations are usually implemented by the Operative System's (OS) kernel. An example of LPWAN is IEEE 802.15.4 (\textit{Low-rate Wireless Personal Area Network})~\cite{ieee_802_15_4} over IPv6, documented as 6LoWPAN in RFC 8930~\cite{6LoWPAN_RFC}, which specifies how to forward 6LoWPAN fragments over a multi-hop network.
The implementation is already available in the Linux Kernel~\cite{6LoWPAN_kernel} for IEEE 802.15.4 compliant devices.
This could benefit \sysname greatly, as it was demonstrated that a proper fragmentation strategy can lead to significant performance improvements~\cite{bruniaux2021defragmenting}.
Yet, not all wireless technologies fit this standard.
For example, there are discrete definitions of header compression and segmentation for LoRa in RFC 9011~\cite{LoRaWAN_compr_frag_RFC}; yet, there are no implementations so far. 

\subsection{Slow TCP Connections}
\label{sec:tech:tcp}

The Transmission Control Protocol (TCP) and how servers handle TCP sessions are usually optimized for fast and reliable connections nowadays, as this is the most common scenario. 
However, for \sysname this is not the case. 
Indeed, in our case, the connections are slow and potentially lossy, leading to problems with many TCP deployments.
The most relevant one is the way retransmissions are handled, as typical deployments are optimized to maximize data rates for fast connections.
One aspect is the Retransmit Timeout (RTO), which is typically set quite low, such that the server can react quickly to network congestion, which is also identified with timeouts.
If the acknowledgement for a packet is not received by the server in time, to optimize data rates for typical connections, the server will quickly do a retransmission.
This results in many duplicate, and thus, unnecessary retransmissions with slow connections called \emph{spurious} retransmissions, quickly saturating the connection.
Another problem to consider is the loss of packets, especially when using one of the wireless communication technologies (cf. \Cref{sec:tech:lpwan}).
For example, today's TCP deployments bundle the transmission of multiple resources over a single connection to achieve a form of concurrency.
However, when a packet is lost for one resource, this delays all of the resources, resulting in even more retransmissions.

Unfortunately, the most effective TCP settings to avoid these issues are controlled by the endpoints, such as retransmission timeouts or the used congestion control algorithm.
Thus, in \sysname, we cannot change these settings, as we can only influence the gateways.
An exception to this is the \texttt{txqueuelen} property for each network interface in Linux.
Settings this to very low values (e.g., 1) prevents the client to send many packets in a short time, which will exacerbate the mentioned retransmission problems.
Another setting to consider is setting TCP's window size, influencing how much data is bundled for each acknowledgment.
Setting this to a low value further helps to avoid retransmission problems.

Another approach to counteract these problems is optimizations for the used wireless communication technologies in a multi-hop setup.
A local retransmission scheme may be used among the gateways, essentially dealing with packet losses on the local gateway network level.
There are different strategies for such local retransmissions~\cite{she2009analytical}.
Handling retransmissions locally would circumvent many of the issues with TCP packet losses.

Furthermore, a more sustainable solution is being deployed at the time of writing and will likely be widely available in the near future.
QUIC~\cite{RFC9000} is a network protocol introduced by Google as a more performant and flexible alternative to TCP. 
Among other functionalities, it leverages UDP with merged handshakes, custom congestion controls, loss detection, and retransmission algorithms~\cite{RFC9002}.
This allows QUIC to handle slow connections with high latency with much better performance than TCP.
For example, Google published a large-scale performance study on QUIC, which showed that QUIC can improve latency by over 30\% compared to TCP with large round-trip-times in a common Internet scenario~\cite{langley2017quic}.
Thus, QUIC would likely have a significant performance impact for our purposes.

Practically speaking, QUIC is already a reality, as many companies are already adopting it.
For example, the company Meta already deployed it for most of its applications~\cite{metaquic}.

\subsection{Changing Routing Paths}
\label{sec:tech:routes}

As described in \Cref{sec:theory}, \sysname needs to reroute packets through different nodes. Typically, TCP sessions stay alive until the end of the communication and it is not possible to dynamically change the stream's endpoints, i.e., the source or destination IP. 
However, changing the output gateway implies a change in the endpoint IP; thus, we need to adopt a strategy for establishing a new endpoint.

A possible way for achieving this goal is using the Reset (RST) flag in the TCP header.
When this flag is set, the receiver will close the TCP session immediately. 
The use of the RST flag is also used for malicious applications, such as the so-called ``TCP Reset Attack" ~\cite{watson2004slipping}.
Here an attacker forges TCP packets with a set RST flag in order to maliciously interrupt or disturb the Internet connection. 
However, for our purposes, the origin gateway can leverage the RST flag to stop a TCP session and force the endpoints (i.e., the client and the Internet server) to start a new session with a different endpoint, i.e., another output gateway selected by the origin.
Indeed, for \sysname, this comes with the additional overhead of establishing a new TCP session regularly.
Thus, when using this approach the \timeout parameter should not be set too low.

However, this is exactly what the Multipath TCP (MPTCP) protocol sets out to do in a more elegant way.
The protocol was originally documented in 2013 by RFC 6824 ~\cite{RFC_multipath_tcp_original} and later updated with RFC 8684 ~\cite{RFC_multipath_tcp}.
This protocol allows a single TCP session to take different routing paths and use different endpoints without interruption.
As widespread support is relevant, MPTCP is already implemented in the Linux Kernel ~\cite{multipath_tcp_kernel} and, e.g., Apple's iOS uses it to be able to quickly switch between a WiFi and cellular connection~\cite{appleMPTCP}.

Another elegant way is provided by the aforementioned QUIC protocol.
While there is also a multipath extension for it in the making~\cite{quicmultipath}, similar to MPTCP, the original QUIC protocol already supports \emph{Connection Migration}~\cite{RFC9000}.
Here, each session is assigned a connection ID, which allows the connection to survive endpoint address changes.

Furthermore, depending on the used underlying protocols for the local network, e.g., IPv4 or LoRaWAN, the client needs additional protection.
Namely, the output gateway should execute a Network Address Translation (NAT) on the connection, translating its own address to the origin.
Otherwise, the address of the origin gateway may leak, and thus, potentially reveal the client's geographic position.

\subsection{Node Discovery}
\label{sec:tech:discovery}

The local gateway network acts as a mesh network, and thus, needs protocols for discovering nodes and establishing routing tables for later network message routing.
For our purposes, a link state routing protocol works well.
While there are many different approaches to these types of protocols, we expect the local gateway network to only observe a limited degree of dynamics, as opposed to, e.g., a mobile ad-hoc network.
The latter typically requires more advanced techniques for routing and node discovery.
Thus, the Optimized Link State Routing Protocol, defined in RFC 3626~\cite{RFC_link_state_routing_v1}, or its successor~\cite{RFC_link_state_routing_v2} is sufficient for our approach.
Here, neighbors exchange information about their neighbors to build simple routing tables, which show the shortest path to any node in the network.

Further, there are also protocols to dynamically optimize the radio settings between nodes.
For example, there is the \emph{Adaptive Data Rate} optimization between LoRa nodes~\cite{kufakunesu2020survey}.
These protocols may be used to further optimize the data rate for the local gateway network.

\subsection{Leakage in Presence of Malicious Nodes}
\label{sec:tech:malicious}
In the mesh and wireless sensor network research area for location privacy (we give an overview of them in \Cref{sec:rw}), there are approaches additionally considering defenses against compromised nodes~\cite{conti2013providing}.
Many approaches like \emph{Network Coding} and \emph{In network location anonymization} are designed to protect against an attacker with a \emph{global} view, which is infeasible in our discussed scenario (cf. \Cref{sec:adv_model}).
These approaches introduce many new requirements and overheads, making them inapplicable for our scenario.
However, approaches considering a \emph{local} adversary are not applicable due to their assumptions.
One type of approach assumes a hierarchical structure among the nodes in the network, in which certain nodes are trusted and cannot be compromised~\cite{el2010hyberloc,el2009hidden,kazatzopoulos2006ihide}.
Due to the nature of our targeted scenario, assuming only some gateways to be trusted is not practical in \sysname.
A different approach is to use end-to-end encryption between source and destination, such that a potentially compromised intermediary node cannot know the endpoints of a message~\cite{sheu2008anonymous}.
However, this requires that all possible endpoint pairs have pre-shared keys deployed, which is infeasible in our scenario.
Another approach can only hide the destination of the connection, not the source~\cite{hong2005effective}.
Yet, protecting the source, i.e., the origin gateway, is the main goal of \sysname.
A different approach protects against compromised nodes, yet relies on the assumption that intermediary nodes cannot be too close to the source~\cite{lightfoot2010preserving}.
We deem this assumption too strict to be practical for \sysname.

The main concern for \sysname regarding compromises is malicious intermediary gateways, which can extract the origin gateway and thus target the client.
Onion routing~\cite{reed1998anonymous} is an effective method to hide the origin gateway from compromised intermediaries. 
No intermediary hop can know its position in the path, as a packet's amount of onion shells is unknown. This is due to \sysname randomly choosing a path length between 0 and \maxhops.
Additionally, onion routing can also be used in wireless networks~\cite{el2018preserving}, where, when mapped to our approach, no node (except for the first and last) is able to learn the source of a message.

\section{Prototype Implementation}
\label{sec:impl}
To demonstrate \sysname, we describe the implementation of our Proof of Concept in this section.
While we have shown advanced techniques in \Cref{sec:tech}, many of them lack either proper wide-ranging support or applicable implementations.
At this point in time, we consider the integration of these techniques as a significant engineering effort and out of scope for this research work.
Note, however, there is potential to significantly improve the practical performance of \sysname.
We will give justifications for the choice of the respective techniques in the following.

\begin{figure}
	\centering
	\includegraphics[width=1\columnwidth,trim={0 0 0 0},clip]{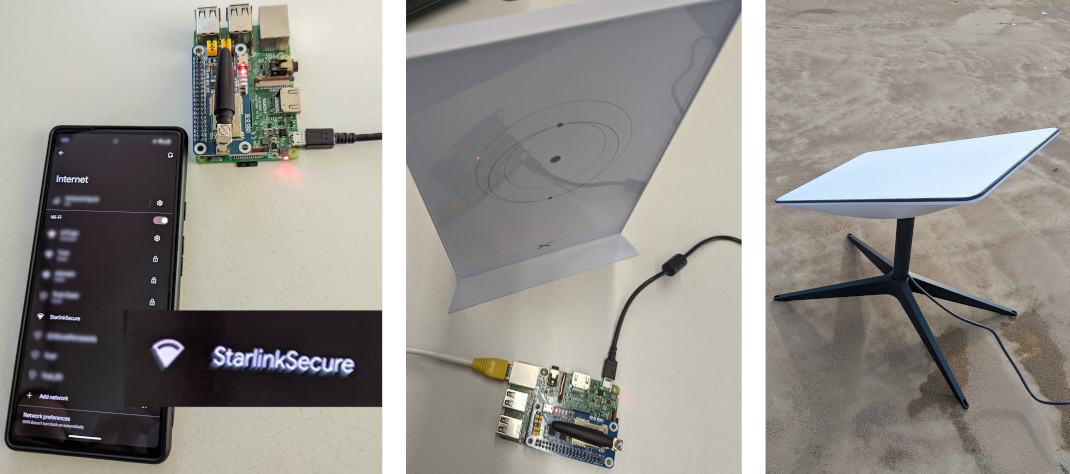}
	\caption{Our test setup. The left picture shows the client with the origin gateway, the center picture shows the output gateway connected to the Starlink router, and the right picture shows the Starlink dish that is connected to the Starlink router.} 
	\label{fig:setup}
\end{figure}

\paragraph{\emph{Scenario}}
For our prototype, we deployed five off-the-shelf consumer devices.
For the Internet connection, we deployed a Starlink dish~\cite{starlinkspecs} on the roof of our building.
We used three gateways deployed as Raspberry Pi 3 Model B+~\cite{rpi3b} (called \emph{RaspberryPi1}, \emph{RaspberryPi2} and \emph{RaspberryPi3}), equipped with a sub-GHz Lora SX1262 868M shield~\cite{lorashield}.
Finally, we used a simple Android phone as the client.
\emph{RaspberryPi3} acts as a proper gateway, as it is connected to the Starlink router, which is connected to the dish.
All Raspberry Pis are interconnected via Lora and provide a WiFi access point for the client to connect.
\Cref{fig:setup} shows the client connected to the origin's secure WiFi access point as well as the output gateway of our test setup.

We have chosen to use sub-GHz Lora due to its use of unlicensed frequency bands.
Another aspect was the availability of development kits that include proper integration for both hardware (e.g., connection to our Raspberries) and software (e.g., driver).
Note, we have also evaluated using the Lora 2.4\,GHz shield SX1280Z3DSFGW1~\cite{lora24kit}; yet, we found that the provided implementation by the manufacturer\footnote{\url{https://github.com/Lora-net/gateway_2g4_hal}} is impractical, as it has limitations that severely limit the effective data rates, even below the Lora sub-GHz shield mentioned above.
For example, the implementation demands a 50\,ms sleep after each packet is sent, meaning a 1500\,bytes MTU split into 7 Lora packets already implies a 350\,ms delay before considering the actual transmission time.

\paragraph{\emph{Implementation}}
To send IP packets over LoRa, we utilize the open source software \texttt{tncattach}\footnote{\url{https://github.com/markqvist/tncattach}} which creates and sets up network interfaces to translate IP to LoRa and back. Unfortunately, the implementation has no support for packet fragmentation and reassembly, and therefore, only utilizes the maximum LoRa packet size (236\,bytes excluding header overhead), leading to the problem discussed in \Cref{sec:tech:mtu}. 
To speed up the translation (and avoid package retransmissions) between Ethernet (typical MTU of 1500\,bytes) and LoRa, we added simple fragmentation support to \texttt{tncattach}.
The more advanced techniques for fragmentation and header compression implemented in the Linux kernel~\cite{6LoWPAN_kernel} are not applicable to LoRa, and the LoRa-specific specification~\cite{LoRaWAN_compr_frag_RFC} has no implementation.

In terms of routing, we pre-deploy our routing tables for simplicity, as opposed to implementing an advanced protocol, as described in \Cref{sec:tech:routes}.
For example, \emph{RaspberryPi1} sets up a Wi-Fi Access Point with its own subnet and forwards all packages to the LoRa interface using another subnet utilizing NAT. \emph{RaspberryPi2} receives these packets and forwards them to \emph{RaspberryPi3}. \emph{RaspberryPi3} is then forwarding them to the Starlink router connected via Ethernet, utilizing NAT again. By providing every device its own static IP address, the operating system takes over tasks such as device discovery and routing.

As the origin gateway, \emph{RaspberryPi1} is in charge of setting up routes for the client's traffic to keep track of the duration of each link. As described in \Cref{sec:tech:routes}, we chose to break TCP connections using a Reset Attack.
We utilized \texttt{scapy}\footnote{\url{https://scapy.net/}} to listen to TCP packets traveling on \emph{RaspberryPi1's} interfaces. As the WiFi Access Point utilizes DHCP, a list of connected devices is known, including the connection time; thus, we can listen to established TCP connections of these devices using a packet filter. If a connection is kept alive for too long (i.e., longer than \timeout), the origin gateway creates a TCP packet with a set RST flag and injects it into the packet flow in both directions, killing the connection. Afterwards, the origin can establish a new route for this client.

Further, to counteract the TCP retransmission problems outlined in \Cref{sec:tech:tcp}, we set the \texttt{txqueuelen} to 1 and TCP's TX buffer to exactly one packet.
This mitigated many spurious retransmissions created because of delayed ACKs.


\section{Evaluation}
\label{sec:eval}
In this section, we evaluate \sysname.
On the one hand, we show real-world results of our prototype setup.
On the other hand, we show the large-scale performance in different settings based on a network simulator run on real-world data sets.

\subsection{Prototype}
\label{sec:eval:prototype}
Recall \Cref{sec:adv_model}, the goal of our approach is to hinder \adv to localize a client, e.g., publishing incriminating information. Usually, such information is published in the form of text or images.
Therefore, to measure the performance of our prototype, we considered three use cases for the client in our setup.
One is simply sending messages via the popular WhatsApp messenger.
Another was to send out tweets via the Twitter Lite app.
The third is to publish an image.
For accurate measurement numbers, we used pings as well as downloaded and uploaded small images via the client.

\paragraph{\emph{Round-Trip-Time (RTT)}} We measured the RTT using a standard ping via zero, one and two LoRa hops to a server on the internet via the Starlink connection. Note that using Starlink alone already adds around 50\,ms of delay. We send 100 pings and averaged the results. The results are depicted in \Cref{tab:ping-eval}.

\begin{table}[]
\centering
\caption{Average RTT and packet loss for pinging 8.8.8.8 with 100 64\,bytes ICMP.}
\begin{tabular}{crc}
\toprule
LoRa Hops &  \multicolumn{1}{c}{RTT}       & Packet Loss \\ 
\midrule
0         & 49.111\,ms   & 0\%         \\
1         & 157.938\,ms & 3\%         \\
2         & 211.786\,ms & 4\%         \\ 
\bottomrule
\end{tabular}
\label{tab:ping-eval}
\vspace{\tblvsp}
\end{table}

\paragraph{\emph{Upload \& Download}} We implemented our own API service to upload images using REST over TLS using HTTP/2. We used a POST form request to upload a picture of size between 50\,kB and 200\,kB. These numbers correspond to the lower and upper bounds of typical mobile applications image compression (e.g., WhatsApp), which we measured with common photographs. Similarly, we downloaded the files using \texttt{wget}. We repeated this experiment 10 times and averaged the results. The results are shown in \Cref{tab:data-eval}.
Note, while we employed both fragmentation for the LoRa packets and optimized the TCP settings, as discussed in \Cref{sec:impl}, we still observed a significant amount of spurious retransmissions.

Naturally, \sysname incurs an overhead. Securely uploading a high-resolution photo using WhatsApp (i.e., 150\,kB) takes 171.92\,s. However, we argue it is reasonable considering the alternative of being either localized or not publishing at all.

\begin{table}[]
\centering
\caption{Average time for uploading and downloading images of different sizes using 2 LoRa hops.}
\begin{tabular}{rrr}
\toprule
\multicolumn{1}{c}{Image Size} & \multicolumn{1}{c}{Upload}  & \multicolumn{1}{c}{Download} \\ 
\midrule
50\,kB  & 65.84\,s  & 60.40\,s   \\
100\,kB & 137.84\,s & 117.80\,s  \\
150\,kB & 171.92\,s & 223.20\,s  \\
200\,kB & 269.31\,s & 276.80\,s  \\ 
\bottomrule
\end{tabular}
\label{tab:data-eval}
\vspace{\tblvsp}
\end{table}

\subsection{Simulation}
\label{sec:eval:sim}
To measure the large-scale performance of \sysname, we implemented a network simulation.
In the following, we describe our evaluation setup by first describing how we simulated the local wireless network, presenting the used real-world data sets, and how the network simulation works.
Afterwards, we present our results showing how the security parameter \maxhops affects the distance from the position of a client, the delays to establish a TLS session in different settings, and finally, the practical data rates.

\subsubsection{Local Wireless Network}
\label{sec:eval:sim:lwn}
While we were restricted to Lora with the sub-GHz frequencies, the simulator allows us to simulate more powerful wireless technologies.
Informed by the data shown in \Cref{tbl:lpwans}, we selected the following combinations of assumed ranges and data rates between nodes:

\begin{enumerate}
    \item 5\,km @50\,kbps (Lora Sub-GHz)
    \item 5\,km @166\,kbps (DASH7)
    \item 1\,km @1\,Mbps (Lora 2.4GHz \& LTE-M Cat-M1)
    \item 1\,km @4\,Mbps (LTE-M Cat-M2)
\end{enumerate}

We excluded NB-IoT due to its low performance compared to the other technologies on our list, which is mostly due to its low-power requirement.
We further excluded Weightless-W, even though its impressive data rates, as we could not establish the readiness of the technology.
For example, unlike the other technologies, we could not find any purchasable devices equipped with Weightless-W components or any practical demonstrators.

Thus, for our simulation we assume, e.g., a range of 5\,km between nodes with a maximum data rate of 50\,kbps, simulating Sub-GHz Lora.
However, the maximum data rate is practically not achievable, especially at longer ranges.
While there are some practical measurements regarding this phenomenon, we found the used evaluation setups (e.g., obstructions or radio settings) and results vary significantly\footnote{For example, one work measured around 10\,kbps~\cite{petajajarvi2017performance} while another measured double the data rate~\cite{swain2021lora} for similar settings.}.
For our simulation, we approximate the \emph{log-distance path loss model} as a simple logarithmic function over the distance between nodes $r=e^{-2d}$. 
$d$ is the distance between two nodes as a relative distance $[0,1]$ with respect to the maximum range.
The resulting data rate $r$ is also relative $[0,1]$ to the maximum data rate.
Thus, the maximum data rate is only achievable if two nodes are right next to each other, while two nodes that are far away, e.g., close to the maximum range, have a severely reduced data rate with only a small fraction of the maximum data rate.

\subsubsection{Data Sets}
\label{sec:eval:sim:dataset}
\begin{figure}
	\centering
	\includegraphics[width=0.32\columnwidth,trim={0 0 0 0},clip]{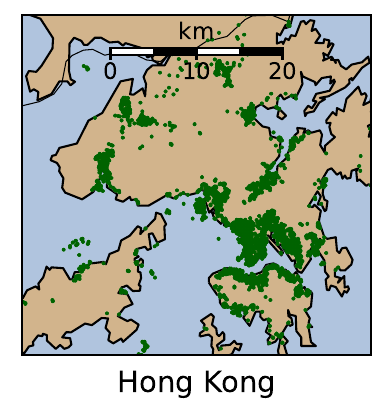}
	\includegraphics[width=0.32\columnwidth,trim={0 0 0 0},clip]{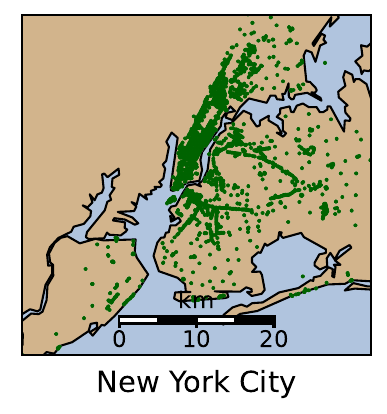}
    \includegraphics[width=0.32\columnwidth,trim={0 0 0 0},clip]{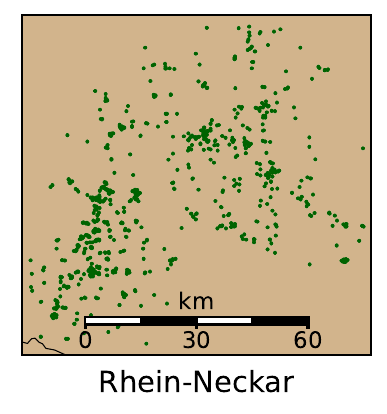}
	\caption{Renderings of the three largest data sets shown in \Cref{tbl:datasets}. Each green dot is one geographic position}.
	\label{fig:datasets}
\end{figure}

As far as we are aware, there are no representative data for gateway positions for the settings we target.
Nevertheless, we found public data sets on various cities' WiFi hotspots to be a good approximation for an urban environment.
\Cref{tbl:datasets} lists all of the data sets we used, each containing the geographic positions of WiFi hotspots for cities of different sizes.
\Cref{fig:datasets} shows renderings of the Hong Kong, New York City, and Rhein-Neckar data sets.

\begin{table}[t]
    \centering
    \caption{Urban WiFi hotspot datasets used for simulation. \emph{Close} refers to the number of records after filtering out too close records. \emph{CC} refers to the number of records contained in the largest connected component when connecting the graph with the given range.}
    \label{tbl:datasets}
    \begin{tabular}{lrrrr}
        \toprule
        Data Set & Total & Close & CC 5\,km & CC 1\,km \\ 
        \midrule
        Hong Kong~\cite{gpshk} & 5441 & 874 & 866 & 332 \\
        New York City~\cite{gpsnyc} & 3319 & 765 & 753 & 602 \\
        Rhein-Neckar~\cite{gpsffrn} & 1338 & 551 & 530 & 30 \\ 
        Brisbane~\cite{gpsbrisbane} & 347 & 97 & 95 & 38 \\
        Paris~\cite{gpsparis} & 277 & 182 & 181 & 178 \\
        Adelaide~\cite{gpsadelaide} & 272 & 51 & 51 & 51 \\
        Leeds~\cite{gpsleeds} & 236 & 169 & 163 & 83 \\
        Linz~\cite{gpslinz} & 124 & 35 & 34 & 29 \\
        \bottomrule
    \end{tabular}
    \vspace{\tblvsp}
\end{table}

However, we had to filter these data sets to fit our needs for the simulation, due to the following two problems.
For one, these data sets contain points that are very close together, which is to be expected, e.g., in the city center, there will be a large number of shops or similar with a high density of hotspots.
As such a dense concentration of gateways is not representative in our settings, we filtered out nodes that are closer than 200\,m from each other.
In \Cref{tbl:datasets}, the number of nodes left after this filtering step is shown as \emph{Close}.
The second problem is that our assumed maximum range leads to a disconnected graph, as some sub-graphs may not be in range for another sub-graph.
Thus, we constructed a graph of nodes from the data sets with the respective maximum range and found the set of connected components in the overall graph.
Finally, we chose the largest connected component for each data set as the actual set of nodes for our simulation.
In \Cref{tbl:datasets}, this is shown as \emph{CC 5\,km} and \emph{CC 1\,km} for a range of 5\,km and 1\,km respectively.
Note, for some data sets the 1\,km range limitation creates a very small network, such as Rhein-Neckar, which covers a large area, but many of the nodes are further than 1\,km away from each other.
Therefore, this results in a significant reduction of the size, if the used connected network, e.g., Rhein-Neckar \emph{CC 1\,km} only has $\sim$5.4\% nodes left relative to \emph{Close}.
However, note that Rhein-Neckar is an exceptional outlier, due to the data set spanning relatively few nodes over almost 100\,km.
The other data sets with higher reductions comprise of dense clusters that are far from one another, e.g., see \Cref{fig:datasets} for Hong Kong.
In such a scenario, \sysname could be applied individually for each cluster.
Generally, a restriction true for all physical wireless networks is that a more densely packed network results in a holistically better connectivity among nodes~\cite{andrew2011computer}.
Therefore, we stress that, if there are no alternatives to using satellite Internet (cf. \Cref{sec:system_model}), even a suboptimal network setup benefits from our approach.

\subsubsection{Network Simulator}
\label{sec:eval:sim:sim}
To implement the network simulation, we used the OMNeT++ 6.0 network simulator~\cite{omnet}.
We load the processed data sets, as described in \Cref{sec:eval:sim:dataset}, as the individual nodes into the simulation.
These nodes are then connected, if they are in range of each other, with a data rate calculated at initialization, as described in \cref{sec:eval:sim:lwn}.
We further use the provided routing component in OMNeT++ to route individual messages as well as to ensure \maxhops is satisfied when randomly selecting gateways by the clients.

For the actual simulation, we assign each client to a random gateway as its origin.
Each client will then proceed to execute the following steps:

\begin{enumerate}
    \item The client's gateway will randomly select an output gateway less than \maxhops hops away.
    \item The client will send a \texttt{TCP} \texttt{SYN} message out to simulate establishing a connection. This message is routed to the output gateway.
    \item After the output gateway received each message, we simulate a 100\,ms delay for the WAN server to answer\footnote{We based this number on the upper average of 50\,ms delay with Starlink and some additional processing time.}.
    \item The output gateway will send a TCP \texttt{SYN-ACK} message back to the client's gateway.
    \item The client will answer this with a TCP \texttt{ACK} and TLS \texttt{ClientHello} combined message, as is a common optimization practice of the Internet.
    \item The simulated server will answer this with a TLS \texttt{ServerHello} message; thus, establishing the TLS connection when the client receives it.
    \item The client will then start sending a 200\,kB data package (the upper bound for WhatsApp image compression) divided into multiple messages, i.e., according to the MTU sizes.
\end{enumerate}

The client will repeat these steps, selecting a new output gateway each time, until the simulation ends after a simulated hour.
Put simply, each client has a fixed origin and will constantly send images with changing output gateways.
For each parameter combination and data set, we execute 30 runs with different random seeds.
While quite a simple setup, this allows us to approximately measure TLS session delays, the effective data rates in the network, and the interactions of both, e.g., TLS session delays of clients over a gateway currently busy sending many data messages.

\subsubsection{Distance to Origin}
\label{sec:eval:sim:dist}
\begin{figure}[t]
	\centering
	\includegraphics[width=\graphsize\columnwidth,trim={0 0 0 0},clip]{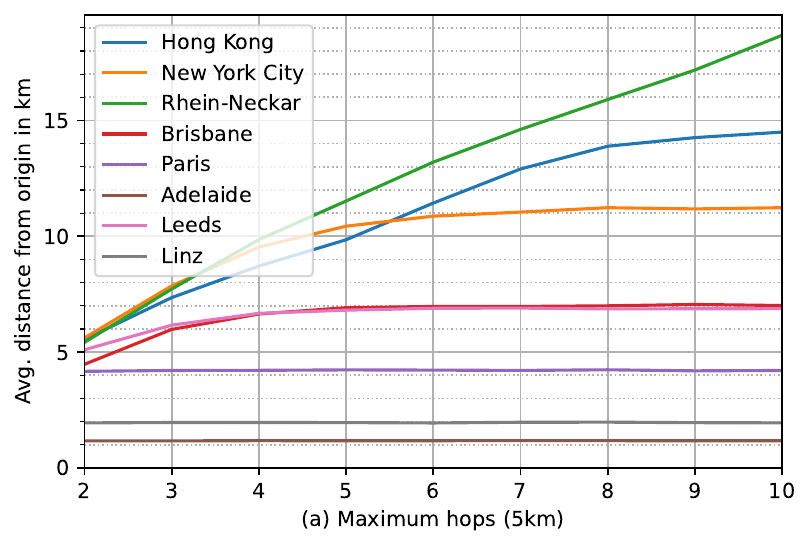}
	\includegraphics[width=\graphsize\columnwidth,trim={0 0 0 0},clip]{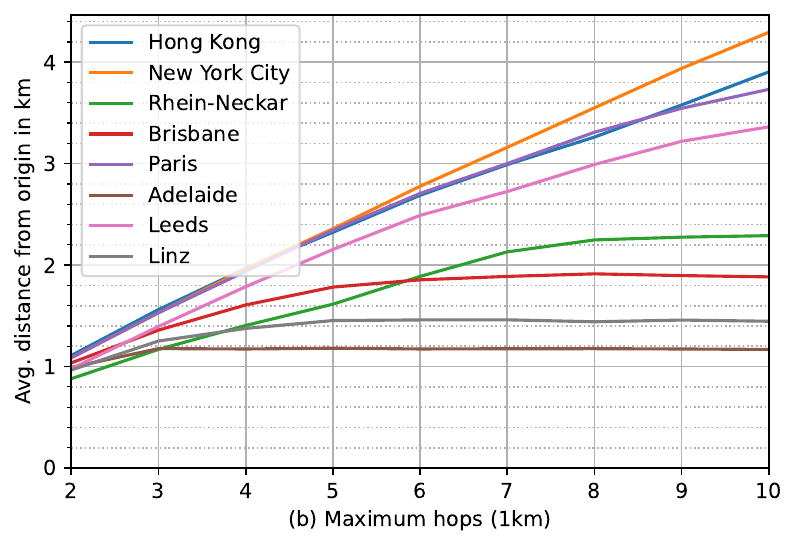}
	\caption{Graphs showing the average distance from a chosen output gateway to the origin for different \maxhops over all data sets with (a) 5\,km and (b) 1\,km range.} 
	\label{fig:avgdist}
\end{figure}

To evaluate the key security parameter \maxhops, we measured the average distance from the client's gateway (origin) to the output gateway.
For this, we employed a simplified simulation over our data sets (cf. \Cref{tbl:datasets}) to get a large sample size of $10\,000$.
For each sample, we select a random origin with a random output gateway less than \maxhops away and measured the actual distance from the origin via their geographic position.
Further, we executed this simulation with differently set \maxhops.
The results of this simulation are shown in \Cref{fig:avgdist}. Wide-ranging data sets in terms of the overall covered area by the nodes, like Hong Kong or Rhein-Neckar, show a nearly linear increase in distance with an increase of \maxhops. We discuss this more thoroughly in \Cref{sec:app:dto}.

\subsubsection{TLS Session Delay}
\label{sec:eval:sim:delay}
To measure the performance of \sysname, we focus on two aspects: delays and data rate.
However, as the primary goal of \sysname is security, we focus on practical applications.
Namely, for delay we measure the average delay it takes to establish a TLS session.
As most Internet services today are based on these sessions, this is a more practical number than measuring simple round-trip delays, especially as we want to see the effects of concurrent data transfers and session establishments in the network.

\begin{figure}[p]
	\centering
	\includegraphics[width=\graphsize\columnwidth,trim={0 0 0 0},clip]{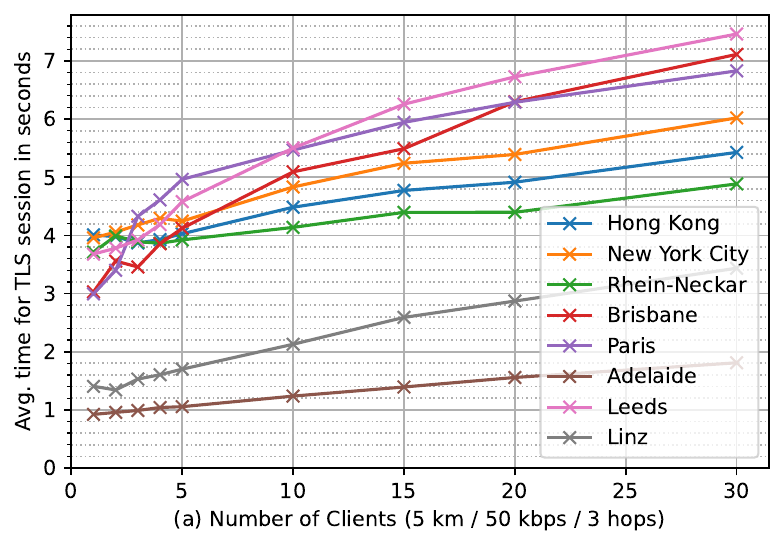}
	\includegraphics[width=\graphsize\columnwidth,trim={0 0 0 0},clip]{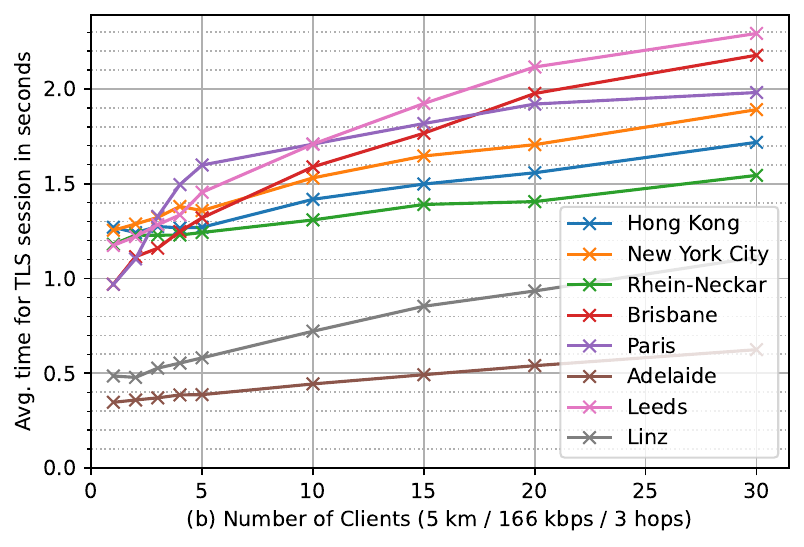}
	\includegraphics[width=\graphsize\columnwidth,trim={0 0 0 0},clip]{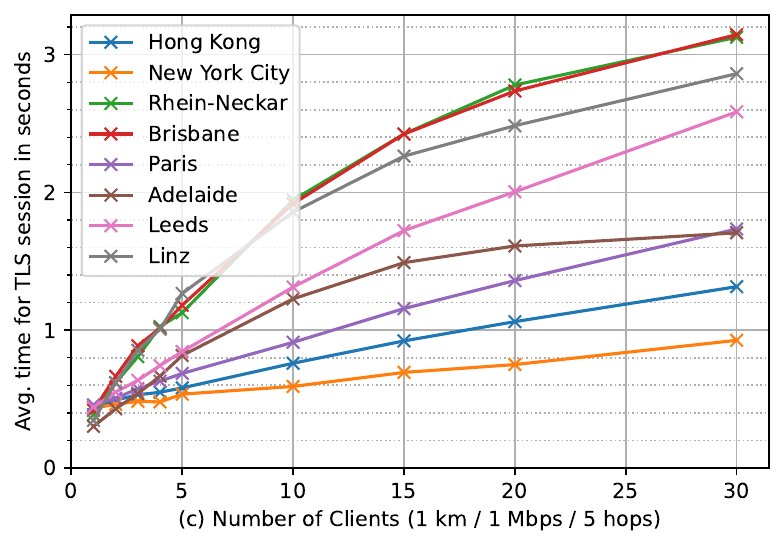}
	\includegraphics[width=\graphsize\columnwidth,trim={0 0 0 0},clip]{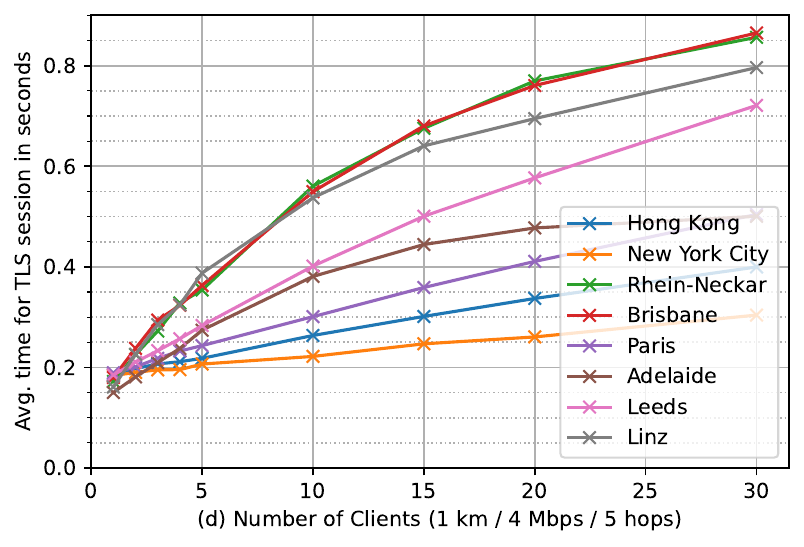}
	\caption{Graphs showing the measurements of TLS session delay in different network types for all data sets.} 
	\label{fig:sim_delay}
\end{figure}

\Cref{fig:sim_delay} shows our measurements.
Noticeable between the slower networks (a) \& (b) and the faster networks (c) \& (d) is the effect of \maxhops with many clients.
With \maxhops = 5 congestion of the gateways affects the delay significantly.
With a range of 5\,km (a) \& (b), the two data sets Adelaide and Linz show a much better performance, due to all nodes being packed closely together in a small area, which also results in much better data rates on average.
Note that there is an implicit tradeoff, as the low average distances between nodes naturally affects the \emph{distance to origin}, as shown in \Cref{sec:eval:sim:dist}.
A similar effect can be observed for the 1\,km range measurements (c) \& (d).
Data sets with closely packed nodes in general, like New York City, show much better performance than data sets with spread-out nodes, like Brisbane.

\subsubsection{Practical Data Rates}
\label{sec:eval:sim:datarate}
In a practical scenario, the data rates may depend on multiple TLS sessions first, which would heavily reduce effective the data rates when simply calculating overall sent bytes divided by time.
Thus, we decided to measure the time it takes to upload a 200\,kB sized image as a practical and isolated example.

\begin{figure}[p]
	\centering
	\includegraphics[width=\graphsize\columnwidth,trim={0 0 0 0},clip]{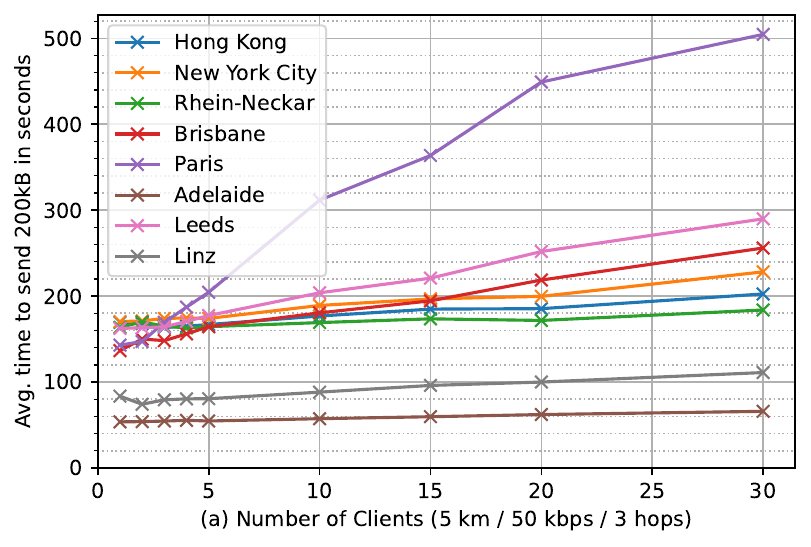}
	\includegraphics[width=\graphsize\columnwidth,trim={0 0 0 0},clip]{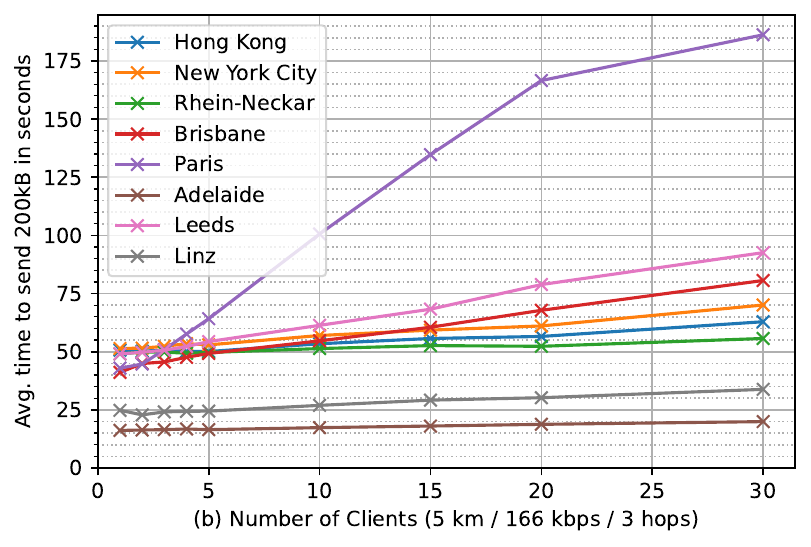}
	\includegraphics[width=\graphsize\columnwidth,trim={0 0 0 0},clip]{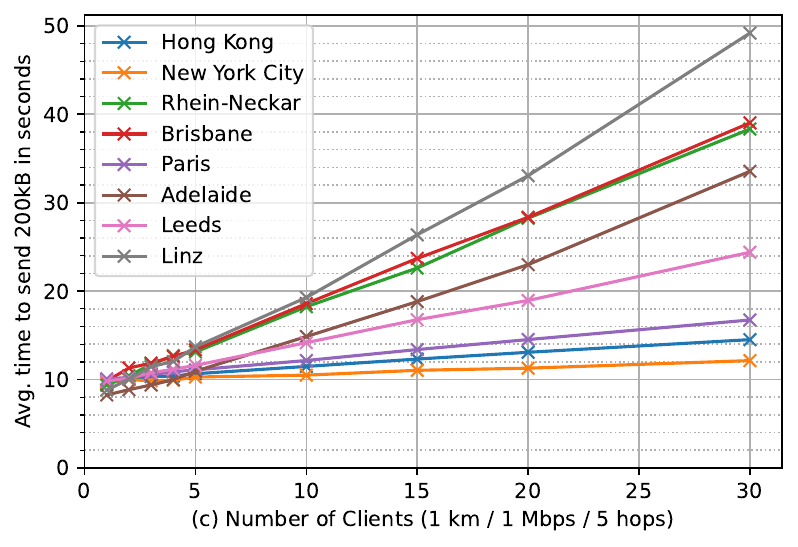}
	\includegraphics[width=\graphsize\columnwidth,trim={0 0 0 0},clip]{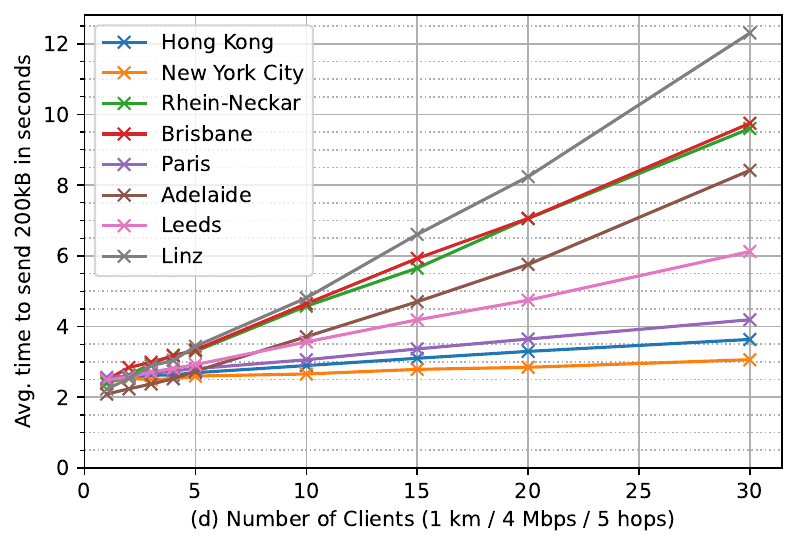}
	\caption{Graphs showing the measurements of length to upload 200\,kB image in different network types for all data sets.} 
	\label{fig:sim_data}
\end{figure}

\Cref{fig:sim_data} shows our measurements.
Generally, the observation made in \Cref{sec:eval:sim:delay} regarding the effect of the density of nodes on data rates applies here as well.
However, while the delays imply a logarithmic trend with a growing number of clients, we can clearly see the effects of many clients sending data and the resulting congestion in the network.
Particularly noticeable are the measurements for Paris with a 5\,km range.
The increase in the number of clients has an especially detrimental effect on the transmission speed.
We believe this is due to the unique and dense spread of nodes in the data set, creating some \emph{bottleneck} nodes that serve multiple connections simultaneously.
This effect disappears with a 1\,km range, as fewer nodes can connect to these bottleneck nodes.


\section{Anonymity \& Security}
\label{sec:security}
In this section, we first analyze the practical anonymity guarantees of \sysname and the security of \sysname.

\subsection{Anonymity}
We quantify and evaluate the anonymity provided by \sysname on the data sets described in \Cref{sec:eval:sim:dataset}.
\Cref{tbl:reach2} summarizes our analysis for each data set for both the 5\,km and 1\,km cases, with  \maxhops = 3 and \maxhops = 5, respectively.
In the following, we will explain our metrics and evaluate our results.

The first metric we evaluated is the traditional \emph{Anonymity Set} size as defined by Chaum~\cite{chaum1988dining}.
The idea is that one's anonymity can be quantified by the set of parties from which one is not distinguishable.
Thus, \adv ideally does not know, which of the parties in the set is the actual target.
In our case, we counted all gateways reachable by an output gateway, i.e., less than \maxhops away, as these are the potential alternatives among the origin gateway.
We considered all gateways in our data sets for both the average and the gateway with the lowest number of reachable gateways in \Cref{tbl:reach2}.
While a dense data set like New York City has a high average, the less dense Rhein-Neckar set has a comparably low average.
When looking at the minima, i.e., the least connected gateways, some data sets contain remote nodes with few reachable nodes.

Nevertheless, just counting the possible alternatives is not a complete metric, as the probabilities to be the origin among the set of reachable gateways may not be uniform.
A way to model this is to consider the entropy of the anonymity set based on implicit information on the network and its nodes leveraged by \adv~\cite{serjantov2003towards}, e.g., the reachability of each node throughout the graph.
In our case, similar to our Anonymity Set metric, we look at all reachable gateways from an output gateway.
In addition, we consider the number of paths between each reachable gateway and the output gateway, as a gateway with more paths to the output gateway is more likely to be the origin than gateways with fewer.
With this entropy, we are able to calculate the \emph{Effective Set} size for the anonymity set (see \Cref{tbl:reach2}).
Concretely, we calculate per reachable gateway $g$ the number of paths to the output gateway divided by all possible paths from any reachable node to the output gateway as $p_g$ (c.f.~\cite{serjantov2003towards}).
With this, we can calculate the effective anonymity set for all $g$ in the reachable gateway set for an output gateway:
\begin{equation*}
    - \sum p_g \log_2 (p_g)
\end{equation*}
We considered all gateways in our data sets to get both the average and the minimum effective anonymity set.
Compared to the uniform anonymity set, data sets that are more clustered in terms of gateway distribution have a significantly lower effective set size (e.g., Rhein-Neckar 5\,km with a ${\sim}39\%$ reduction) than data sets that are well-connected (e.g., Paris 5\,km with a ${\sim}4\%$ reduction).
These numbers demonstrate that different data sets with different degrees of connectivity among the gateways result in varying degrees of anonymity, which needs to be considered in practical deployment.
However, the average effective anonymity set sizes show the efficacy of \sysname.

Finally, we measured the average number of paths between any two gateways as \emph{Node2node Paths} within \maxhops distance.
This gives a hint at the general connectivity among the nodes.
To further refine this metric, we also counted all \emph{Unique Paths}, i.e., the number of paths that do not share any common gateways in their route.
This indicates how reliable \sysname is when individual gateways fail, i.e., the number of alternative paths.
In \Cref{tbl:reach2}, we can examine that in more dense networks with many connections, like Paris 5\,km, there are many alternative paths on average.
Contrarily, if there are few connections between nodes, like Paris 1\,km, then there are few alternative paths on average.
Generally, a higher number shows a more resilient network against gateway failures.

\begin{table*}[t]
    \centering
    \caption{Anonymity metrics for all data sets for both the 5\,km and 1\,km cases.}
    \label{tbl:reach2}
    \newcommand\rot{45}
    \newcommand\ol{-26pt}
   \begin{tabular}{rrrrrrrrrrrrr}
         & \rotatebox{\rot}{\parbox{\widthof{Anonymity Set}}    {5\,km Average \\Anonymity Set}} \hspace{\ol} & 
         \rotatebox{\rot}{\parbox{\widthof{5\,km Minimum}}      {5\,km Minimum \\Anonymity Set}} \hspace{\ol} &
         \rotatebox{\rot}{\parbox{\widthof{5\,km Average}}      {5\,km Average \\Effective Set}} \hspace{\ol} &
         \rotatebox{\rot}{\parbox{\widthof{5\,km Minimum}}      {5\,km Minimum \\Effective Set}} \hspace{\ol} &
         \rotatebox{\rot}{\parbox{\widthof{Node2node Paths}}    {5\,km Average \\Node2node Paths}} \hspace{\ol} &
         \rotatebox{\rot}{\parbox{\widthof{5\,km Average}}      {5\,km Average \\Unique Paths}} \hspace{\ol} &
         \rotatebox{\rot}{\parbox{\widthof{Anonymity Set}}      {1\,km Average \\Anonymity Set}} \hspace{\ol} & 
         \rotatebox{\rot}{\parbox{\widthof{5\,km Minimum}}      {1\,km Minimum \\Anonymity Set}} \hspace{\ol} &
         \rotatebox{\rot}{\parbox{\widthof{5\,km Average}}      {1\,km Average \\Effective Set}} \hspace{\ol} &
         \rotatebox{\rot}{\parbox{\widthof{5\,km Minimum}}      {1\,km Minimum \\Effective Set}} \hspace{\ol} &
         \rotatebox{\rot}{\parbox{\widthof{Node2node Paths}}    {1\,km Average \\Node2node Paths}} \hspace{\ol} &
         \rotatebox{\rot}{\parbox{\widthof{5\,km Average}}      {1\,km Average \\Unique Paths}} \hspace{-6pt} \\
        \midrule
        Hong Kong       & 417.3 & 12 & 278.0 & 9.1 & 6\,080.3 & 54.2
                        & 111.8 & 26 & 60.1 & 21.8 & 5\,147.9 & 5.7 \hspace{18pt} \\
        New York City   & 523.7 & 13 & 330.0 & 12.6 & 6\,415.1 & 59.9
                        & 92.7 & 8 & 47.7 & 8.0 & 3\,734.4 & 4.2 \hspace{18pt} \\
        Rhein-Neckar    & 98.5 & 11 & 60.4 & 7.5 & 193.1 & 8.9
                        & 18.3 & 13 & 12.2 & 9.1 & 116.2 & 2.3 \hspace{18pt} \\
        Brisbane        & 83.7 & 17 & 61.9 & 14.9 & 862.9 & 20.2
                        & 35.3 & 30 & 26.7 & 20.8 & 116.5 & 4.3 \hspace{18pt} \\
        Paris           & 180.0 & 179 & 172.7 & 162.4 & 10\,096.9 & 93.8
                        & 78.4 & 12 & 43.0 & 9.4 & 332.2 & 2.7 \hspace{18pt} \\
        Adelaide        & 50.0 & 50 & 50.0 & 50.0 & 2\,402.0 & 50.0
                        & 50.0 & 50 & 44.7 & 43.7 & 2\,045.5 & 13.4 \hspace{18pt} \\
        Leeds           & 149.7 & 49 & 121.4 & 39.5 & 1\,880.1 & 35.6
                        & 37.9 & 10 & 22.5 & 8.4 & 61.2 & 2.1 \hspace{18pt} \\
        Linz            & 33.0 & 33 & 33.0 & 33.0 & 959.8 & 31.6
                        & 37.9 & 26 & 20.7 & 19.2 & 834.3 & 4.6 \hspace{18pt} \\
        \bottomrule
    \end{tabular}
    \vspace{\tblvsp}
\end{table*}

\subsection{Security}
The adversary \adv aims to geolocate a specific client by identifying the origin gateway.
\adv may use the following strategies to accomplish this: (1) \adv assumes the used output gateway is close enough and target it instead, (2) \adv uses individual intercepted messages from the local network to trace the origin, and (3) \adv monitors the gateways for extended periods to collect data pointing to the origin.

Strategy (1) is prevented in our system by rerouting communication away from the origin gateway used by the client.
In \Cref{sec:eval:sim:dist}, we analyze the effective distances achieved on real-world data sets.
Nevertheless, due to our \emph{Changing Gateway} approach (cf. \Cref{sec:overview}), eventually, the actual origin gateway is used for the satellite uplink.
Thus, \adv may simply target each gateway and eventually be successful.
However, this would effectively result in a large-scale attack, e.g., in an urban context, targeting the entire city.
As stated in our adversary model (\Cref{sec:adv_model}), we deem this undesirable for \adv.
Note that this also applies to \adv taking kinetic measures against gateways in subregions of the network. As described in \Cref{tbl:reach2} (column \emph{Unique Paths}), every network has at least two unique paths between any two nodes, i.e., alternative paths that do not share any nodes.
This demonstrates the network's resilience against forceful disconnection, even in the presence of a costly, wide-ranging attack on many gateways instead of targeting individual ones.
We deem this undesirable, as \adv tries to limit its attacks as much as possible.
Further, in case \adv captures extensive territory, resulting in few gateways in the region, we assume that clients will not remain in the area and transmit sensitive data.

For the second strategy (2), \adv is prevented from learning any information about the route with the end-to-end encryption employed by the gateways.
Yet, \adv may intercept numerous local gateway messages over time and eventually be able to correlate the route from the output gateway back to the origin.
Our system prevents this by regularly changing the output gateway (cf. \Cref{sec:overview}).
The effectiveness of this approach is dependent on the gateway distribution, which we evaluate in \Cref{sec:eval:sim:dist} on our real-world data sets.

\adv may monitor all used output gateways by a client to infer the origin.
The success of this strategy (3) is prevented by the \emph{Selection Bias}, as described in \Cref{sec:overview}, which shifts the centroid of all used gateways over time away from the origin.

In the following, we discuss additional local attacks, motivating our adversary model.

\paragraph{\emph{Total Local Network Monitoring}}
In \Cref{sec:adv_model}, we assume \adv is unable to monitor the entire gateway network to establish a holistic view of the local network.
\adv would aim to trace back routes from the output gateway back to the origin.
\adv would need to get an extensive coverage of the gateway network.
Practically speaking, to achieve this, \adv needs to deploy numerous devices, which must be widely spread in close proximity to the gateways and potentially deployed over long periods, as it is unknown where and when the client may become active.
Considering the potential scale of an active conflict, we deem this strategy infeasible.

\paragraph{\emph{Jamming}}
\adv may try to use jamming to interrupt the client's ability to communicate.
There are three types of jamming to consider.
One type is targeting the satellites, e.g., with a high-power ground-based jamming signal directed at individual satellites.
However, with over 3\,000 Starlink satellites deployed~\cite{Istarlinktotalno} at the time of writing, this strategy does not scale well.
Further, according to reports, Starlink used specialized firmware updates to withstand numerous jamming attacks~\cite{Istarlinkupdate}.

An additional type of jamming is adversarial satellites jamming ground stations.
Theoretically, a signal can be considered jammed if the Signal to Noise Ratio (SNR) at the receiver is 1.
To achieve this the jamming signal must be received with at least the same power as the benign signal~\cite{mpitziopoulos2009effective}.
The power of electromagnetic signals decays quadratically with distance.
Thus, a satellite targeting a ground station would need a jamming signal powerful enough to cover the vast distances in space.
As satellites are significantly limited in terms of power, we deem this strategy infeasible.

Another strategy is local jamming ground-to-ground, as modern jammers may cover a wide area.
However, depending on the scale of the gateway network, \adv would need to either deploy many jammers or target the client's general area, which might be unknown.
In case \adv \emph{is} able to jam the client's gateway, this is difficult to circumvent; however, in this case, \adv is not able to geolocate the client.

\paragraph{\emph{Malicious Gateways}}
\adv may try to deploy malicious gateways.
However, we deem this strategy to be unviable.
Similar to the holistic monitoring of the gateway network, \adv would need to deploy or compromise a plethora of devices spread throughout the gateway network to reliably trace clients.
Individual gateways may reveal the client's route if the client actually routes over them.
However, unlike \emph{overlay} networks, in which an adversary can remotely create new nodes, \adv must control \emph{physical} gateways in advantageous geographic locations to reliably target clients.
Thus, such an elaborate strategy requires physical access, resulting in low probabilities of success.
To protect against leakage of the origin gateway by intermediary malicious nodes, approaches such as onion routing can be deployed to \sysname, as described in \Cref{sec:tech:malicious}.


\section{Related Work}
\label{sec:rw}

To the best of our knowledge, our approach is the first work to address the triangulation of satellite Internet users in critical circumstances.
Thus, we could not find directly related work for our purposes.
However, one related research topic is providing \emph{location privacy} in mesh and wireless sensor networks.
Another related research topic is \emph{emergency communication networks}, which, similarly to \sysname, often employ mesh-like network topologies and are designed to work under exceptional situations.

\paragraph{\emph{Location Privacy in Mesh and Wireless Sensor Networks}}
In these networks, due to their physical properties, communication is vulnerable to tracking and monitoring.
If vulnerable participants utilize such a network, keeping the (geographic) position undisclosed becomes essential.
Approaches for location privacy can be grouped into eleven categories~\cite{conti2013providing}.

The high-level idea of \emph{Random Walk}-based approaches is to direct packets to traverse a network through a random path to a sink (e.g., base station).
This makes the path of a packet unpredictable to an adversary attempting local traffic analysis~\cite{kamat2005enhancing, xi2006preserving}.
\emph{Geographic Routing} is similar to Random Walk, yet utilizes the physical location information of nodes for more efficient routing of packets towards the sink.
For location privacy, these approaches additionally leverage pseudonyms, reputation, and a fixed set of intermediary nodes creating a mix subnetwork~\cite{shaikh2010achieving, li2009preserving}.
Both these types of approaches assume a \emph{backtrack} or \emph{hunter} adversary model, in which the adversary starts close to the sink, traces each sent-out message back to the next hop, and this is repeated until the adversary eventually arrives at the source.
However, this is incompatible with our assumptions, as all nodes in \sysname are sinks, we regularly switch the sink, and a backtracking adversary is unlikely in a conflict zone.

In \emph{Delay}-based solutions, nodes store incoming packets and transmit them after a random period of time, disrupting the chronological order of the packets. 
This also modifies the traffic pattern, rendering it difficult for a local adversary to trace the origin of the traffic~\cite{hong2005effective, kamat2009temporal}.
\emph{Limiting node detectability} temporarily throttles or disables the transmission power of nodes, making it harder for an adversary to receive packets~\cite{rios2011exploiting, el2010hyberloc}.
\emph{Network Coding} uses homomorphic encryption at intermediary nodes to hide traffic flows. 
After receiving the aggregated data, the sink is then able to reverse the encryption process~\cite{fan2009efficient, fan2010preventing}.
However, these three classes of solutions add significant delays to the network that adversely affect low-latency networks, such as \sysname, especially when the aim is to be compatible with the Internet (cf., \Cref{sec:tech:tcp}).

\emph{Dummy Data Sources} generate authentic-looking dummy traffic to obscure the authentic traffic.
The objective is to prevent an adversary from differentiating between genuine and fabricated traffic~\cite{yang2013towards, mehta2007location}.
\emph{Cyclic Entrapment} confuses potential adversaries by routing the traffic between nodes with cyclical patterns~\cite{ouyang2006entrapping, kazatzopoulos2006ihide}.
\emph{Separate Path Routing} splits data into multiple packets, which will be sent over multiple, non-intersecting paths to the sink. 
Therefore, the local adversary is only able to capture part of the data~\cite{wang2009privacy}.
Similarly to introducing delays, these three approaches induce large traffic overheads to the network (e.g., dummy traffic or additional retransmissions). 
Thus, they are incompatible with the goals of \sysname (cf., \Cref{sec:requirements}).

\emph{Cross Layer Routing} utilizes multiple OSI layers to hide information from adversaries~\cite{shao2009cross}.
Yet, this is based on the assumption that the adversary may not see certain OSI layers, which we deem impractical.
Approaches focusing on \emph{Wireless Mesh Networks} specifically, usually assume a hierarchical network with base stations, mesh routers, and mesh clients.
Numerous security features, including location privacy, rely on pseudonyms in the form of public key certificates, either directly distributed by an authority~\cite{zhang2006arsa} or are self-generated with a domain authority backing it~\cite{sun2010sat}.
\emph{In network location anonymization} utilizes hierarchical (e.g., clusters of nodes) pseudonyms or aggregation of traffic to hide the source of the traffic~\cite{misra2006efficient, luo2010location}.
Both the Wireless Mesh Networks and In network location anonymization approaches assume a hierarchical trust structure in the network, which is not feasible in \sysname's local network as, e.g., all gateways are set up by citizens and trusted equally.

\paragraph{\emph{Emergency Communication Networks}}%
These networks are designed to provide reliable communication during an emergency when other communication infrastructures fail. 
The research community has since proposed many approaches for establishing a network in case of natural disasters, conflicts, or any adverse situation that prevents typical access to the Internet.
In particular, Portmann \etal~\cite{portmann2008wireless} defined the fundamental characteristics an emergency network must have in its design: \emph{privacy, data integrity, authentication, and access control}. 
The most common emergency networks are considering the use of Locally Deployed Resource Units (LDRU) (e.g., base stations of cellular networks), satellites, ad-hoc networks, or a combination of them.
Usually, portable devices span a network, while only a subset of them are actually capable of external means of communication (e.g., satellites)~\cite{pradeep2015survey, kishorbhai2017aon}.

To establish emergency networks, many recent works focus on using Unmanned Aerial Vehicles (UAVs)~\cite{debnath2021comprehensive}.
Numerous works in this area leverage \emph{drones} to establish the emergency network.
One proposal is to directly leverage the drones as base stations~\cite{zhao2019uav}.
Other works use drones to span a mesh network back to a static base station.
Proposed wireless communication technologies for the mesh range from leveraging WiFi~\cite{panda2019design}, LTE~\cite{deruyck2018designing}, 5G~\cite{gao2020intelligent}, and LoRa~\cite{pan2021uav}.
Besides the use of drones, the application of aerostatic balloons found space in the context of emergency networks. 
One approach is to build an ad-hoc network based on IEEE 802.11j between the ballons~\cite{shibata2009disaster}.
Another approach is to span a multihop WiFi backbone from one area to a satellite-based base station via zeppelin-like balloons~\cite{suzuki2006ad}, while another extends this approach to a mesh network~\cite{okada2012network}.
Besides all the possible proposed designs, multiple recent works are proposing several optimizations, including load balancing between the units~\cite{lin2021adaptive, niu20213d}.

A practical concern is that drones exhibit limited fly times, and thus, coverage. Further, the individual hardware required (i.e., the drones and balloons) is expensive; yet, hundreds or even thousands of units are necessary to operate in an emergency. Moreover, drones and aerostatic balloons are subject to weather conditions (e.g., cannot easily fly in a storm), which is heavily limiting their operative scenario.

Other works focus on deploying an ad-hoc mesh network between base stations that are then put into communication with satellites. 
Zhou \etal~\cite{zhou2021integrated} develop such a network based on WiFi 2.4\,GHz for the network backbone and 5.8\,GHz for data transmission. However, due to the limited range of WiFi, this approach requires the devices to be quite close to each other and does not scale well to cover a wide area. 
Instead, Iapichino \etal~\cite{iapichino2008advanced} propose a hybrid system where equipped vehicles (Vehicle Communication Gateways) establish a connection with satellites and users can connect to these mobile gateways. 
Similarly, Patricelli \etal~\cite{patricelli2009disaster} are proposing a MOBSAT access point (that has to be carried by car or helicopter), which provides high-speed data connection to the users through WLAN and WiMAX through GEO satellites.
Nevertheless, these approaches assume that numerous, specially equipped vehicles are prepared and ready for use.

In contrast, the target scenario of \sysname is novel in the context of emergency networks.
The continuous expansion of satellite Internet services allows for the deployment of comparably cheap access points.
This enables to provide widespread access to many deployed satellite base stations.
Therefore, the mentioned works above are not taking the unique problems of this scenario into account.
As outlined in the Introduction, \sysname specifically aims to protect its users from triangulation.
Furthermore, our focus is on the accessibility of the design system.
Thus, \sysname does not require any additional application or hardware installed on the user's device and the gateways are easy to deploy.


\section{Conclusion}
\label{sec:conclusion}
In this work, we presented \sysname, the first scheme to address the triangulation of satellite Internet users, and thus, protect them from being targeted.
To achieve this, \sysname leverages a local wireless communication technology to span a network between the gateways, rerouting a client's connection away from the origin gateway.
Additionally, \sysname regularly changes the output gateway for each client to avoid detection of long-lasting connections.
We thoroughly discussed different technical aspects, meeting our defined requirements to make \sysname usable with both existing Internet protocols and widely available hardware.
We implemented a prototype demonstrating \sysname's feasibility and evaluated its performance via network simulation on various real-world data sets.

\section*{Acknowledgment}
This work was supported by the European Space Operations Centre with the Networking/Partnering Initiative.

\bibliographystyle{IEEEtran}
\bibliography{IEEEabrv,bib}

\newpage

\appendices
\section{Distance To Origin Evaluation}\label{sec:app:dto}
The data sets with an assumed 5\,km maximum range between nodes in \Cref{fig:avgdist} (a) shows the limitations of the different data sets.
In contrast, very densely populated sets that do not cover a lot of area, like Adelaide or Paris, show no increase in distance.
This effect is primarily dependent on the set's covered area.
For example, if we roughly draw a circle around the nodes provided by each data set, we get a diameter of $\sim$5\,km for Adelaide and $\sim$15\,km for Paris, while Hong Kong has $\sim$50\,km and Rhein-Neckar even $\sim$100\,km.
Data sets, like New York City ($\sim$30\,km) or Brisbane ($\sim$20\,km) that cover a medium-sized area, show a diminishing return in terms of an increased \maxhops.
We also found that some sets have more uniformly spread nodes over the covered area, while others have a decreasing density from the center to the borders of the covered area.
For example, the more uniformly dense Paris reaches the limit sooner than the unevenly dense Brisbane.
With the Adelaide data set, we cannot get past 5\,km and when considering the random selection of nodes with a more dense center, our average distance from the origin is naturally quite low.

\Cref{fig:avgdist} (b) shows similar measurements for the 1\,km range case.
We can see the effects analogously to the 5\,km results, just with a significantly reduced distance from the origin.
Note that some data sets, especially Rhein-Neckar, have severely reduced node count when considering a connected network with a 1\,km range (cf. \Cref{sec:eval:sim:dataset}).
The effect of this can be seen in this graph.
Overall, there is an inherent trade-off for setting the \maxhops parameter, as setting it higher leads to an increased distance; yet, more hops for the communication will lead to more delays and overhead in the network.
Our measurements show that it is crucial to take the underlying spread and density of nodes into account when choosing \maxhops.
For the following simulation results, we assumed \maxhops = 3 for the 5\,km range networks and \maxhops = 5 for the 1\,km range networks.

\section{Result Graphs}\label{sec:app:rg}

\clearpage

\end{document}